\documentclass[aps,prd,superscriptaddress,nofootinbib,preprintnumbers,floatfix,tightenlines,fleqn,notitlepage,10pt]{revtex4-1}
\usepackage{amsmath,amsfonts,amssymb,amscd,amsxtra,amsthm}
\usepackage{graphicx}  
\usepackage{epstopdf}
\usepackage{dcolumn}  
\usepackage{bm}          
\usepackage{slashed}
\usepackage{cancel}
\usepackage{float}
\usepackage{mathtools} 
\usepackage{amsbsy}
\usepackage{amstext}

\usepackage[utf8]{inputenc}
\usepackage{booktabs}
\usepackage[normalem]{ulem}
\usepackage[dvipsnames]{xcolor}
\usepackage{tabularx}
\usepackage{enumitem}
\usepackage{array}
\usepackage{slashed}
\usepackage{tikz}
\usepackage{pgfplots}
\pgfplotsset{compat=1.18}  
\usepackage{float}
\usepackage{multirow}
\usepackage{cancel}
\usepackage{physics}
\usepackage{hyperref}
\usepackage{cleveref}

\renewcommand\sout{\bgroup \color{red} \ULdepth=-.5ex \ULset}

\begin{document}
\preprint{INHA-NTG-05/2025}
\title{Production mechanism of doubly charmed exotic mesons $T_{cc}$} 

\author{Hee-Jin Kim}
\affiliation{Department of Physics, Inha University, Incheon 22212,
  Republic of Korea} 
\author{Hyun-Chul Kim}
\affiliation{Department of Physics, Inha University, Incheon 22212,
  Republic of Korea}
\affiliation{Physics Research Institute, Inha University, Incheon
  22212, Republic of Korea}
\affiliation{School of Physics, Korea Institute for Advanced Study
(KIAS), Seoul 02455, Republic of Korea}
\date{\today}

\begin{abstract}
We investigate the production mechanism for doubly charmed tetraquark
mesons within a coupled-channel formalism. The two-body Feynman kernel
amplitudes are constructed using effective Lagrangians that respect
heavy quark symmetry, chiral symmetry, SU(3) flavor symmetry, and
hidden local symmetry. The fully off-shell coupled scattering
equations are solved within the Blankenbecler-Sugar (BbS) reduction
scheme. We find three positive-parity and one negative-parity
tetraquark states with total spin $J=1$. Among them, two
positive-parity states appear as bound states in the isoscalar and
isovector $DD^*$ channels, while another appears as a resonance in the
$D^*D^*$ channel. A negative-parity resonance is also predicted in the
isoscalar channel. We analyze the coupling strengths of these
tetraquark states to various channels. The dependence of the results
on the reduced cutoff mass $\Lambda_0$ is examined. The most
significant tetraquark state remains stable within the range of
$\Lambda_0=(600-700)$ MeV. 
\end{abstract}
\maketitle

\section{Introduction}
Since the Belle Collaboration announced the existence of the
$X(3872)$, which is also known as 
$\chi_{c1}(3872)$~\cite{Belle:2003nnu}, the understanding of 
hidden-charm tetraquark states has undergone remarkable development 
both experimentally and theoretically (see the following reviews and 
references therein~\cite{Lebed:2016hpi, Guo:2017jvc, Guo:2019twa,
  Yamaguchi:2019vea, Chen:2022asf, Meng:2022ozq}).   
In 2022, the LHCb Collaboration reported the discovery of the first 
doubly charmed tetraquark state, $T_{cc}(3875)^+$, observed as a 
narrow enhancement near the $D^0D^{*+}$ mass 
threshold~\cite{LHCb:2021vvq}. Its Breit-Wigner mass was found to
lie just below the $D^0D^{*+}$ threshold 
with a remarkably narrow width of $(410 \pm 164)$~keV. The quantum
numbers are tentatively assigned as $I(J^P) = 0(1^+)$, though an
isovector assignment has not been definitively excluded by current
data. Notably, approximately 90\% of the observed $D^0D^0\pi^+$ events
in the energy region between the $D^0D^{*+}$ threshold and $3.9$~GeV
contain a genuine $D^{*+}$ meson~\cite{LHCb:2021auc}. This recent
finding of the $T_{cc}(3875)^+$ tetraquark state has triggered a great
deal of theoretical work (see recent reviews and references
therein~\cite{Chen:2022asf, Meng:2022ozq}). Even before the
observation of the $T_{cc}(3875)^+$, the existence of doubly charmed
tetraquark states had already been speculated and
investigated~\cite{Ballot:1983iv, 
  Zouzou:1986qh, Kim:1995bm, Pepin:1996id, Moinester:1995fk,
  Ebert:2007rn, Navarra:2007yw, Ohkoda:2012hv, Eichten:2017ffp,
  Luo:2017eub, Zhu:2019iwm, Tang:2019nwv, Tan:2020ldi, Cheng:2020wxa}.    
Theoretical studies can largely be divided into two categories: one
considers the $T_{cc}$ as a compact tetraquark state composed of four
quarks~\cite{Cheng:2020wxa, Praszalowicz:2022sqx, Zhang:2022qtp,
  Kucab:2024nkv,  Anwar:2023svj, Meng:2023for, Dong:2024upa},  
while the other views it as a molecular state of two $D$ ($D^*$)
mesons~\cite{Ren:2021dsi, Albaladejo:2021vln, Dai:2021wxi, Du:2021zzh, 
  Ling:2021bir, Chen:2021cfl, Feijoo:2021ppq, Fleming:2021wmk,
  Ke:2021rxd, Xin:2021wcr, Xin:2021wcr, Zhao:2021cvg, Agaev:2022ast,
  Kamiya:2022thy, Meng:2022ozq, Asanuma:2023atv, Vidana:2023olz,
  Abolnikov:2024key, Abolnikov:2024key, Meng:2024kkp, Whyte:2024ihh,
  Wu:2024zbx, Lu:2025zae}. Lattice QCD 
studies~\cite{Padmanath:2022cvl,Lyu:2023xro,Meng:2024kkp} 
have demonstrated the presence of a pole in $DD^*$ scattering near the
physical point and have explored the isovector channel, suggesting possible
additional exotic states in the doubly charmed sector.

Coupled-channel formalisms provide a very useful framework to analyze
the production mechanisms of hadron resonances, particularly exotic
ones (see, for example, recent reviews~\cite{Mai:2022eur,
  Doring:2025sgb}). The dynamical generation of the $T_{cc}(3875)^+$
was also well described within several coupled-channel
approaches~\cite{Albaladejo:2021vln, 
  Dai:2021wxi, Du:2021zzh, Kamiya:2022thy, Vidana:2023olz}.   
Among the various coupled-channel approaches, we employ a formalism
based on two-body Feynman kernel amplitudes, developed by
the Bonn-J\"ulich group~\cite{Machleidt:1987hj,
  Haidenbauer:2005zh}. This approach has been successfully applied to 
a wide range of hadronic reactions, including
$\pi\pi$~\cite{Lohse:1990ew}, $\pi\rho$~\cite{Janssen:1993nj},
$\pi\eta$~\cite{Janssen1995}, $\pi N$~\cite{Schutz:1998jx},
$NN$~\cite{Kim:1994ce,Janssen:1996kx}, and
hyperon-nucleon~\cite{Reuber:1995vc, Haidenbauer:2005zh} interactions.   
Whereas the Bonn-J\"ulich group employed time-ordered perturbation
theory, we implement the Blanckenbecler-Sugar (BbS)
scheme~\cite{Blankenbecler:1965gx,Aaron:1968aoz}, which is manifestly
Lorentz-covariant. We also adopt a more constrained treatment of the
cutoff masses in the hadronic form factors to reduce uncertainties.

This coupled-channel formalism was successfully applied to
the $\pi\rho$, $\pi\omega$, and $K\bar{K}^*$ coupled
interactions~\cite{Clymton:2022jmv, Clymton:2023txd, Clymton:2024pql}, where 
the production mechanisms and internal structures of axial-vector mesons
were explained: the $a_1$ meson can be interpreted as a
$K\bar{K}^*$ molecular state~\cite{Clymton:2022jmv}, and $b_1$ mesons
exhibit two-pole structures~\cite{Clymton:2023txd}. Furthermore, this
approach clarified the discrepancies among existing
experimental data for the $h_1$ mesons~\cite{Clymton:2024pql}. The
framework was extended to the heavy-light sector, successfully
describing the $D_{s0}^*(2317)$ as a $DK$ molecular
state~\cite{Kim:2023htt}. It was also applied to explain how the
hidden-charm pentaquark states $P_{c\bar{c}}$, observed by the LHCb
Collaboration~\cite{LHCb:2015yax, LHCb:2019kea,
  LHCb:2021chn}, can be generated from baryon–heavy meson
interactions~\cite{Clymton:2024fbf}. Moreover, it provided a theoretical
explanation for the absence of such pentaquark resonances in
$J/\psi$ photoproduction observed by the GlueX
Collaboration~\cite{GlueX:2019mkq, GlueX:2023pev}. The production
mechanisms of hidden-charm pentaquark states with strangeness
$S = -1$, reported by the LHC and Belle \& Belle II
Collaborations~\cite{LHCb:2020jpq, LHCb:2022ogu, Belle:2025pey}, were
also described within this coupled-channel
formalism~\cite{Clymton:2025hez}, which was further extended to
predict double-strangeness hidden-charm pentaquark
states~\cite{Clymton:2025zer}. 

\begin{figure}
  \centering
  \includegraphics[scale=1.0]{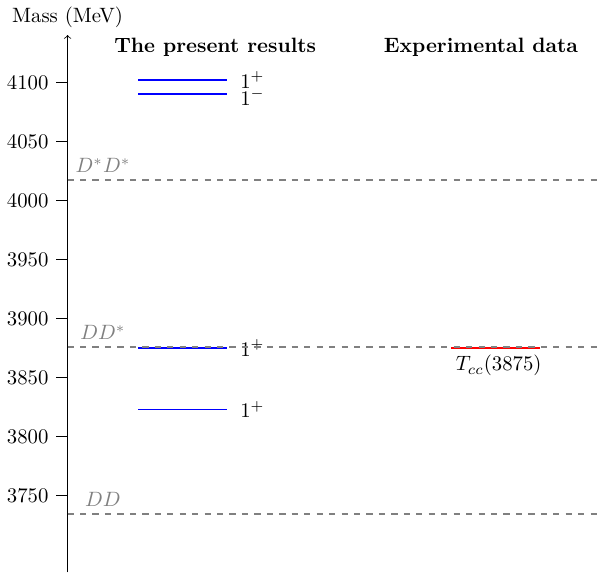}
  \caption{Mass spectrum of $T_{cc}$ states in the present work.
  The spin-parity quantum numbers are written for each state.
  The red line denotes the observed $T_{cc}(3875)$ by the LHCb
  Collaboration.}
  \label{fig:1}
\end{figure}

In this work, we investigate the production mechanism for doubly
charmed mesons within a coupled-channel formalism. We construct the
two-body Feynman kernel amplitudes using effective Lagrangians that
respect heavy quark symmetry, chiral symmetry, SU(3) flavor
symmetry, and hidden local symmetry. To solve the fully off-shell
coupled scattering integral equations, we adopt the
Blanckenbecler-Sugar (BbS) scheme, a three-dimensional reduction of
the Bethe-Salpeter equation. The dynamically generated
resonances appear as either poles on the real axis or in the
complex plane of the transition amplitudes. We characterize these
tetraquark states by analyzing their pole positions and coupling
strengths to the involved channels.
We summarize the current results in Fig.~\ref{fig:1}, which will be
discussed in detail.

The structure of the present work is outlined as follows: In Section
II, we explain the off-shell coupled-channel formalism. In Section
III, we analyze the kernel amplitudes for various channels,
demonstrate the emergence of the $T_{cc}(3875)^+$ state as a $DD^*$
molecular state in the isovector--axial-vector channel by examining
the partial-wave transition amplitudes with both positive and negative
parities, and predict the existence of three exotic states with
$I(J^P)=0(1^+)$, $1(1^+)$, and $0(1^-)$. We dicuss their feature by
extracting pole positions and coupling strengths and uncertainty
asrising from parameters. Finally, in Section IV, we summarize the key
findings of the current investigation, and draw conclusions.

\section{Theoretical Framework}
\begin{figure}[htbp]
  \centering
  \includegraphics[scale=1.0]{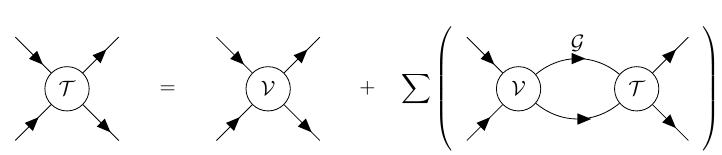}
  \caption{Graphical representation of the coupled  Bethe-Salpeter
    integral scattering equation.}
  \label{fig:2}
\end{figure}

We start by defining the scattering amplitude as 
\begin{align} 
S_{fi} = \delta_{fi} - (2\pi)^4 \delta^{(4)}(P_f-P_i) \mathcal{T}_{fi},
\label{eq:1}
\end{align} 
where $P_{i}$ and $P_f$ represent the total four-momenta of
the initial and final hadron states, respectively. The transition
amplitudes $\mathcal{T}_{fi}$ contains essential 
information on dynamics of hadronic interactions.
They are derived by solving the Bethe-Salpeter (BS) equation 
\begin{align}
\mathcal{T}_{fi}(p,p';s) = \mathcal{V}_{fi}(p,p';s) + \frac{1}{(2\pi)^4}
  \sum_k \int d^4q\, \mathcal{V}_{fk}(p,q;s) G_k(q;s),
  \mathcal{T}_{ki}(q,p';s)
  \label{eq:2} 
\end{align}
where $p$ and $p'$ denote the relative four-momenta of the intial and
final states, respectively. $s$ stands for the square of the total
cnergy in the center-of-mass (cm) frame: $s=P_i^2=P_f^2$.  
$q$ is the off-mass-shell momentum for the intermediate states in the
cm frame. $\mathcal{V}_{fi}$ indicates the two-body Feynman kernel
amplitudes, which can be constructed by using the effective
Lagrangian. The summation runs over intermediate coupled channels
involved in the process $f\to i$. The two-body propagator $G_i$ in the
off-shell intermediate states is expressed as: 
\begin{align}
G_i = \frac{i}{q_{i1}^2-m_{i1}^2}\frac{i}{q_{i2}^2-m_{i2}^2}.
\label{eq:3}
\end{align}
The coupled BS integral equations given in Eq.~\eqref{eq:2} are
illustrated in Fig.~\ref{fig:2}. 

Since it is complicated to deal with the four-dimensional BS equation
numerically, we will perform a three-dimensional reduction, following 
the Blankenbecler-Sugar (BbS) scheme~\cite{Blankenbecler:1965gx,
  Aaron:1968aoz} by implementing a specific form of the two-body
propagator:   
\begin{align}
G_k(q) = \frac{\pi}{\omega_1^k \omega_2^k} \delta\left(q^0 - 
\frac{\omega_1^k-\omega_2^k}{2}\right) 
\frac{\omega_1^k+\omega_2^k}{s-(\omega_1^k+\omega_2^k)^2+i\epsilon}, 
\label{eq:4}
\end{align}
where $\omega_{1(2)}^k=\sqrt{\bm{q}^2+(m_{1(2)}^k)^2}$ denotes the
on-shell energies for particles 1 and 2 in channel $k$. Thus, we
derive the three-dimensional coupled BbS integral equation expressed
as 
\begin{align}
T_{fi}(\bm{p},\bm{p'};s)=V_{fi}(\bm{p},\bm{p'};s)+ \sum_k \int d^3q \, 
  V_{fk}(\bm{p},\bm{q};s) \tilde{G}_k(\bm{q}) T_{ki}(\bm{q}, \bm{p'};s)
\label{eq:5}  
\end{align}
with
\begin{align}
\tilde{G}_k(\bm{q}) = \frac{1}{(2\pi)^3} 
\frac{\omega_1^k+\omega_2^k}{2\omega_1^k\omega_2^k} 
\frac{1}{s-(\omega_1^k+\omega_2^k)^2+i\epsilon}.
\label{eq:6}  
\end{align}

To clarify the spin-parity assignments for $T_{cc}$ states, we
project the $\mathcal{V}$ and $\mathcal{T}$ amplitudes into the
partial-wave amplitudes, which results in a one-dimensional coupled
partial-wave BbS integral equation:  
\begin{align}
T_{\lambda'\lambda}^{J(fi)} (\mathrm{p},\mathrm{p}';s) &=
  V_{\lambda'\lambda}^{J(fi)}(\mathrm{p},\mathrm{p}';s) +
  \sum_{\lambda_k} \frac{1}{(2\pi)^3} \int_0^\infty dq\, 
\frac{\omega_1^k+\omega_2^k}{2\omega_1^k\omega_2^k}
\frac{\mathrm{q}^2V_{\lambda'\lambda_k}^{J(fk)}
 (\mathrm{p},\mathrm{q};s) T_{\lambda'\lambda_k}^{J(ki)} 
(\mathrm{q},\mathrm{p}';s)}{s-(\omega_1^k+\omega_2^k)^2+i\epsilon}, 
\label{eq:7}
\end{align}
where the partial-wave kernel and transition amplitudes $V^J$ and
$T^J$ are defined by
\begin{align}
V_{\lambda'\lambda}^{J(fi)}(\mathrm{p},\mathrm{p}';s) &= 2\pi \int
 d\cos\theta 
 V_{fi}(\mathrm{p},\mathrm{p}',\theta;s)  d_{\lambda\lambda'}^J(\theta),  
\cr
T_{\lambda'\lambda}^{J(fi)} (\mathrm{p},\mathrm{p}';s) &= 2\pi \int
  d\cos\theta T_{fi}(\mathrm{p},\mathrm{p}', \theta;s)
 d_{\lambda\lambda'}^J(\theta).
\label{eq:8}  
\end{align}
Here, $d_{\lambda\lambda'}^J(\cos\theta)$ denotes the reduced Wigner $D$
function and $\theta$ is the scattering angle between $\bm{p}$ and 
$\bm{p}'$. Here, $\lambda=\lambda_1-\lambda_2$ and
$\lambda'=\lambda_1'-\lambda_2'$ indicate the helicity differences
between particles 1 and 2 in the initial and final states, 
respectively. $\mathrm{p}$, $\mathrm{p'}$, and $\mathrm{q}$ designate
the magnitude of the corresponding three-momenta.
We separate the singularity in Eq.~\eqref{eq:27} arising from the
two-body propagator as follows:
\begin{align}
  \mathcal{T}^{J(fi)}_{\lambda'\lambda} (\mathrm{p}',\mathrm{p}) = 
  \mathcal{V}^{J(fi)}_{
  \lambda'\lambda} (\mathrm{p}',\mathrm{p}) + \frac{1}{(2\pi)^3}
  \sum_{k,\lambda_k}\left[\int_0^{\infty}d\mathrm{q}
  \frac{\mathrm{q}E_k}{E_{k1}E_{k2}}\frac{\mathcal{F}(\mathrm{q})
  -\mathcal{F}(\tilde{\mathrm{q}}_k)}{s-E_k^2}+ \frac{1}{2\sqrt{s}}
  \left(\ln\left|\frac{\sqrt{s}-E_k^{\mathrm{thr}}}{\sqrt{s}
  +E_k^{\mathrm{thr}}}\right|-i\pi\right)\mathcal{F}
  (\tilde{\mathrm{q}}_k)\right]
\label{eq:9}
\end{align}
with 
\begin{align}
\mathcal{F}(\mathrm{q})=\frac{\mathrm{q}}{2}\,
  V^{J(fk)}_{\lambda'\lambda_k}(\mathrm{p}',
  \mathrm{q})T^{J(ki)}_{\lambda_k\lambda}(\mathrm{q},\mathrm{p}) ,
\label{eq:10}  
\end{align}
and $\tilde{\mathrm{q}}_k$ denotes the momentum $\mathrm{q}$ when
$E_{k1}+E_{k2}=\sqrt{s}$.

To construct the two-body Feynman kernel amplitudes, we employ the 
effective Lagrangians for $D$ and $D^*$ mesons, considering
heavy-quark spin-flavor symmetry, SU(3) flavor symmetry, chiral
symmetry, and hidden local symmetry. 
In the $m_Q\to \infty$ limit, the effective Lagrangian can be
written as 
\begin{align}
  \label{eq:11}
\mathcal{L}_\mathrm{heavy}^Q &= ig\mathrm{Tr}
[H_b \gamma_\mu\gamma_5 \mathcal{A}_{ba}^\mu\bar{H}_a] 
+ i\beta\mathrm{tr}[H_b v_\mu(\mathcal{V}-\rho)_{ba}^\mu 
\bar{H}_a] + i\lambda\mathrm{tr}[H_b \sigma_{\mu\nu}
F_{ba}^{\mu\nu} \bar{H}_a] + g_\sigma H_a H_a \sigma,
\end{align}
where $H_a$ denotes a superfield consisting of the pseudoscalar and
vector heavy meson fields $P$ and $P_\mu^*$
\begin{align}
\label{eq:12}
H_a = \frac{1+\slashed{v}}{2} \left[\slashed{P}_a^*
-P_a\gamma_5\right]
\end{align}
in $4\times 4$ Dirac space. The superfield preserves the heavy quark
spin-flavor symmetry and satisfies $\slashed{v}H=H$. The limit $m_Q\to
\infty$ requires the field normalization to be
\begin{align}
  \label{eq:13}
\langle 0|P_a|(Q\bar{q})_a,0^-\rangle = \sqrt{M_H}, \;\;\;
\langle 0|P_a^{*\mu}|(Q\bar{q})_a,1^-\rangle = \epsilon^\mu 
\sqrt{M_H},
\end{align}
with the polarization vector of the vector heavy meson,
$\epsilon^\mu$. The conjugate field is then expressed as 
\begin{align}
  \label{eq:14}
\bar{H}_a = \gamma^0 H_a^\dagger \gamma^0 = 
\left[\slashed{P}_a^{*\dagger}
+P_a^{\dagger}\gamma_5\right] \frac{1+\slashed{v}}{2}
\end{align}
The heavy quark flavor symmetry allows $P^{(*)}$ to be either $P^{
(*)}=\{D^{(*)0}(c\bar{u}),D^{(*)+}(c\bar{d}),D_s^{(*)+}(c\bar{s})\}$
for $Q=c$ or $P^{(*)}=\{B^{(*)-}(b\bar{u}),B^{(*)0}(b\bar{d}),B_s^ {(*)0}(b
\bar{s})\}$ for $Q=b$.

The matter fields $H_a$ interact with Goldstone
bosons via the chiral transformation $H \to HU^\dagger$. The
Goldstone fields appear as the coset field 
$\xi$ of $\mathrm{SU (3)}_L\times \mathrm{SU(3)}_R/\mathrm{SU
(3)}_V$~\cite{Coleman:1969sm,Callan:1969sn}:
\begin{align}
  \label{eq:15}
\xi(x) = e^{i\mathcal{M}(x)/f_\pi}, \;\;\; \text{with} \;\;\;
\mathcal{M} = \left(
\begin{array}{ccc}
\frac{\pi^0}{\sqrt{2}}+\frac{\eta}{\sqrt{6}} & \pi^+ & K^+
\\ 
\pi^- & -\frac{\pi^0}{\sqrt{2}}+\frac{\eta}{\sqrt{6}} & K^0
\\
K^- & \bar{K}^0 & -\sqrt{\frac{2}{3}}\eta
\end{array}\right),
\end{align}
where $f_\pi=132\,\mathrm{MeV}$ is the pion decay constant.
The axial-vector and vector currents are then constructed from 
the $\xi$ field:
\begin{align}
  \label{eq:16}
\mathcal{A}^\mu &= \frac1{2}(\xi^\dagger
\partial^\mu\xi-\xi\partial^\mu\xi^\dagger) = \frac{i}{f_\pi}
\partial^\mu\mathcal{M} + \cdots, \cr
\mathcal{V}^\mu &= \frac1{2}(\xi^\dagger
\partial^\mu\xi+\xi\partial^\mu\xi^\dagger) = \frac{i}{2f_\pi^2}
[\mathcal{M},\partial^\mu\mathcal{M}] + \cdots,
\end{align}
which transform under $U$ as
\begin{align}
  \label{eq:17}
\mathcal{A}^\mu &\to U\mathcal{A}^\mu U^\dagger, \cr
\mathcal{V}^\mu &\to U\mathcal{V}^\mu U^\dagger
-\partial^\mu UU^\dagger.
\end{align}
Thus,  the effective chiral Lagrangians for
the $\mathcal{M} PP^*$ and $\mathcal{M}P^*P^*$ vertices can be
constructed as follows 
\begin{align}
\mathcal{L}_{PP^*\mathcal{M}} ={}& -\frac{2g}{f_\pi} P_a^{Q\dagger} 
\partial^\mu \mathcal{M}_{ba} P_{b\mu}^{Q^*} + \mbox{h.c.}, \cr
\mathcal{L}_{P^*P^*\mathcal{M}} =& - \frac{2ig}{f_\pi} 
\varepsilon^{\mu\nu\alpha\beta} P_{a\alpha}^{Q*\dagger} v_\beta 
\partial_\mu \mathcal{M}_{ba} P_{b\nu}^{Q*}.
\label{eq:18}
\end{align}

The interaction for the vector meson octet is established via hidden
local symmetry. A gauge equivalence between $\mathrm{SU (3)}_L\times 
\mathrm{SU(3)}_R/\mathrm{SU(3)}_V$ and $\mathrm{SU (3)}_L\times 
\mathrm{SU(3)}_R \times [\mathrm{SU(3)}_V]_\mathrm{local}$ allows the
emergence of the vector mesons as dynamical gauge bosons of hidden
local symmetry~\cite{Bando:1984ej,Bando:1985rf}. The vector meson
octet is then given by 
\begin{align} 
\rho^\mu = i\frac{g_V}{\sqrt{2}} V^\mu,\;\;\;
V^\mu=\left(\begin{array}{ccc}
\frac{\rho^0}{\sqrt{2}}+\frac{\omega}{\sqrt{2}} & \rho^+ & K^{*+}
\\ 
\rho^- & -\frac{\rho^0}{\sqrt{2}}+\frac{\omega}{\sqrt{2}} & K^{*0}
\\
K^{*-} & \bar{K}^{*0} & \phi
\end{array}\right)^\mu.
\label{eq:19}
\end{align}
The corresponding field strength tensor is defined as $F^{\mu\nu} =
\partial^\mu \rho^\nu - \partial^\nu \rho^\mu + [\rho^\mu,\rho^\nu]$, 
which gives rise to the tensor interaction term in
Eq.~\eqref{eq:11}. Therefore, the effective Lagrangians for 
the $PPV$, $P^*P^*V$, and $PP^*V$ vertices are expressed as 
\begin{align}
\mathcal{L}_{PPV} =& - \sqrt{2} \beta g_V\, P_a^{\dagger} v_\mu 
V_{ba}^\mu P_b, \cr
\mathcal{L}_{P^*P^*V} =& \sqrt{2} \beta g_V\, P_{a\nu}^{*\dagger} 
v_\mu V_{ba}^\mu P_b^{*\nu} + 2\sqrt{2}i\lambda g_V\, 
P_{a\mu}^{*\dagger} P_{b\nu}^{*} V_{ba}^{\mu\nu} , \cr
\mathcal{L}_{PP^*V} =& -2\sqrt{2}\lambda g_V\, 
\varepsilon_{\beta\alpha\mu\nu} \left(P_a^{\dagger} v^\beta 
P_b^{*\alpha} + P_a^{*\dagger\alpha} v^\beta P_b\right) 
\left(\partial^\mu V^\nu\right)_{ba},
\label{eq:20}
\end{align}
where the heavy quark velocity can be replaced by
$i\overleftrightarrow{\partial}_\mu/(2\sqrt{MM'}) =
i(\overleftarrow{\partial}-\overrightarrow{\partial})/(2\sqrt{MM'})$,
with $M^{(')}$ denoting the mass of the heavy meson. 
The effective Lagrangians for the scalar meson $\sigma$ are
obtained as 
\begin{align}
\mathcal{L}_{PP\sigma} =& - 2g_{PP\sigma} \, P_a^{\dagger} P_a \,
\sigma, \cr
\mathcal{L}_{P^*P^*\sigma} =& 2g_{P^*P^*\sigma} \, P_{a\mu}^
                              {*\dagger} P_a^{*\mu} \, \sigma.
\label{eq:21}  
\end{align}

We now proceed to determine the coupling constants in the effective
Lagrangians. The value of $g$ in Eq.~\eqref{eq:18} is fixed by the
partial decay width of $D^{*+}\to D^0\pi^+$~\cite{CLEO:2001foe}: 
\begin{align}
\Gamma(D^{*+} \to D^0\pi^+) = \frac{g^2}{6\pi f_\pi^2} |\bm{p}_\pi|^3,  
\label{eq:22}
\end{align}
which yields $g=0.59$. The value of $\beta=0.9$ in Eq.~\eqref{eq:20}
is determined by using the vector meson
dominance~\cite{Bando:1984ej}. The value of $\beta=0.9$ in
Eq.~\eqref{eq:20} is determined by using vector meson
dominance~\cite{Bando:1984ej}, and $\lambda=0.56\,\mathrm{GeV^{-1}}$
is fixed phenomenologically by the $B\to K^*$ transition form
factor at high $q^2$~\cite{Casalbuoni:1996pg, Isola:2003fh}.     

For the $\sigma PP$ and $\sigma P^*P^*$ coupling constants, the
$\sigma$ meson was often assumed to be a chiral partner of the pion,
leading to the relation $g_\sigma = g_\pi/(2\sqrt{6})$. Moreover, it
was assumed that $g_{\sigma P^*P^*}$ equals $g_{\sigma PP}$. However,
due to the large width of the $\sigma$ meson (also known as
$f_0(500)$)~\cite{PDG:2024cfk}, it is more plausible to consider the
$\sigma$-exchange as a correlated $2\pi$-exchange in $S$-wave. There
is no fundamental reason to require $g_{\sigma PP}=g_{\sigma
  P^*P^*}$. In Ref.~\cite{Kim:2019rud}, the coupling constants
$g_{\sigma PP}$ and $g_{\sigma P^*P^*}$ were determined by using
dispersion relations within this theoretical framework. For the
charmed mesons, we obtained the following results: $g_{\sigma
  DD}=1.50$ and $g_{\sigma D^*D^*}=5.21$. Since the 
$\rho$-meson exchange is regarded as correlated $2\pi$-exchange in
$P$-wave, we similarly determine $g_{\rho DD}=1.65$ and $g_{\rho
  D^*D^*}=6.47$. The $\rho DD^*$ coupling constant is fixed at
$g_{\rho DD^*} =5.8$ using the relation $g_V=m_\rho/f_\pi$ from the
KSRF relation~\cite{Kawarabayashi:1966kd, Riazuddin:1966sw}.  

\begin{figure}[!htp]
\centering
\includegraphics[scale=.8]{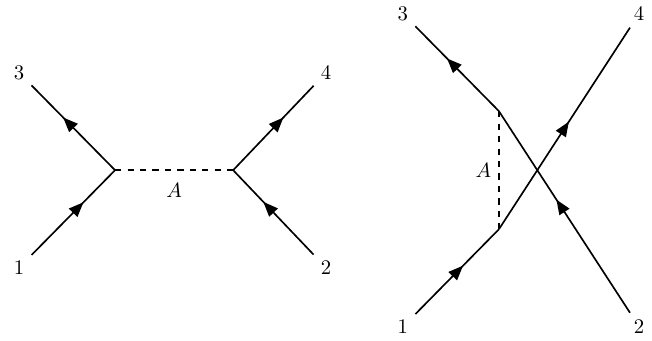}
\caption{$t$- and $u$-channel Feynman diagrams.}
\label{fig:3}
\end{figure}

Since the mass of $T_{cc}(3875)^+$ is found to be below the
$D^0D^{*+}$ threshold, we consider three channels, namely, $DD$,
$DD^*$, and $D^*D^*$. The kernel matrices are therefore constructed
as: 
\begin{align}
\mathcal{V} = 
\left(\begin{array}{@{}ccc@{}}
\mathcal{V}_{DD\to DD} & 
\mathcal{V}_{DD\to DD^\ast} &
\mathcal{V}_{DD\to D^\ast D^\ast} \\
\mathcal{V}_{DD^\ast\to DD} & 
\mathcal{V}_{DD^\ast\to DD^\ast} &
\mathcal{V}_{DD^\ast\to D^\ast D^\ast} \\
\mathcal{V}_{D^\ast D^\ast\to DD} & 
\mathcal{V}_{D^\ast D^\ast\to DD^\ast} &
\mathcal{V}_{D^\ast D^\ast\to D^\ast D^\ast} \\
\end{array}
\right),
\label{eq:23}
\end{align}
where each matrix element is expressed in terms of the initial and
final momenta, $\mathrm{p}$ and $\mathrm{p}'$. The kernel amplitudes
are derived as the sum of the Feynman amplitudes in the $t$- and
$u$-channels at tree level: 
\begin{align}
  \label{eq:24}
\mathcal{V}_{i\to j} = \sum_A \mathcal{M}_{i\to j}^A,
\end{align}
where $A$ represents all possible particle exchanges for the process
$i\to j$. Note that $s$-channel pole diagrams are not included since
we focus on the dynamical generation of resonances. Figure~\ref{fig:3}
depicts generic diagrams for the $t$- and $u$-channels.  

The Feynman kernel amplitude for the exchange of a particle $A$ has
the structure  
\begin{align}
  \label{eq:25}
\mathcal{M}_{i\to j}^A=\mathrm{IS} F_A^2 \Gamma_1^A \mathcal{P}^A
\Gamma_2^A,
\end{align}
where $\mathrm{IS}$ denotes the isospin factor derived from the SU(3)
Clebsch-Gordan coefficients. In this expression, $\Gamma_1^A$
($\Gamma_2^A$) is the vertex function for the initial (final) state,
and $\mathcal{P}^A$ is the propagator of the exchanged particle $A$. A
form factor $F_A$ is introduced at each vertex to account for the
finite size of the corresponding hadron and to ensure the unitarity of
the transition amplitudes. We choose the form factor as  
\begin{align}
  \label{eq:26}
F_A(q^2) = \left(\frac{n\Lambda^2-m_A^2}
{n\Lambda^2-q^2}\right)^n,
\end{align}
where $\Lambda$ is the cutoff mass, $m_A$ is the mass of
the exchanged particle, and $q$ denotes the momentum transfer between
the incoming and outgoing states. In the limit $n\to \infty$, $F_A$
reduces to a Gaussian form. In this work, we fix $n=2$, since the
results are not sensitive to this choice. 

The cutoff masses are the main source of uncertainty because they
cannot be precisely determined due to the lack of experimental
data. To minimize these uncertainties, we adopt the idea that heavier
hadrons are more compact than lighter ones~\cite{Kim:2018nqf,
  Kim:2021xpp, Won:2022cyy}, so that the cutoff mass can be regarded
as the inverse of the particle size. Consequently, it increases
monotonically with the mass of the exchanged particle. Based on this,
we define the reduced cutoff mass $\Lambda_0$ as $\Lambda_0 \equiv
\Lambda - m_{\mathrm{ex}} \approx (600-700)$ MeV.
This strategy has the benefit of eliminating the need to fit each
cutoff mass individually. This approach has been phenomenologically
successful in describing various hadronic processes~\cite{Kim:2019rud,
  Clymton:2022jmv, Clymton:2023txd}, particularly those involving
heavy hadrons~\cite{Clymton:2023txd, Kim:2019rud, Kim:2023htt}. In the
present work, we fix the reduced cutoff mass to $\Lambda_0 = 600$ MeV
and examine how the doubly charmed tetraquark resonances arise. Note
that this value is not fitted to the existing data on
$T_{cc}(3875)$. The uncertainty due to $\Lambda_0$ is discussed by
varying it within $\pm 100$ MeV in Section III. The values of the
various $\mathrm{IS}$ factors and $\Lambda_0$ for the isoscalar and
isovector channels are listed in Tables~\ref{tab:1} and \ref{tab:2},
respectively.   
\setlength{\tabcolsep}{15pt}
\renewcommand{\arraystretch}{1.5}
\begin{table}[!htp]
\centering
\caption{Values of the IS factors and reduced cutoff masses for
each vertex and isospin factor for the kernel amplitudes for $DD$,
$DD^\ast$ and $D^*D^*$ in the isoscalar($I=0$) channel.}
\label{tab:1}
\begin{tabular}{c|c|c|c|c}
\hline
\hline
\multirow{2}{*}{Reaction} & \multirow{2}{*}{Exchange} & 
\multicolumn{2}{c|}{IS factor} & 
\multirow{2}{*}{$\Lambda_0$ [MeV]} \\
\cline{3-4}
& & $t$ channel & $u$ channel \\
\hline
\multirow[t]{5}{*}{$DD\to DD$} 
& $\sigma$ & $1$ & $-1$ & 600 \\
& $\rho$ & $-3/2$ & $3/2$ & 600 \\
& $\omega$ & $1/2$ & $-1/2$ & 600 \\
\multirow[t]{5}{*}{$DD\to DD^*$} 
& $\rho$ & $-3/2$ & $3/2$ & 600 \\
& $\omega$ & $1/2$ & $-1/2$ & 600 \\
\multirow[t]{5}{*}{$DD\to D^*D^*$} 
& $\pi$ & - & $3/2$ & 600 \\
& $\eta$ & - & $-1/6$ & 600 \\
& $\rho$ & $-3/2$ & $3/2$ & 600 \\
& $\omega$ & $1/2$ & $-1/2$ & 600 \\
\hline
\multirow[t]{5}{*}{$DD^*\to DD^*$} 
& $\sigma$ & $1$ & - & 600 \\
& $\pi$ & - & $3/2$ & 600 \\
& $\eta$ & - & $-1/6$ & 600 \\
& $\rho$ & $-3/2$ & $3/2$ & 600 \\
& $\omega$ & $1/2$ & $-1/2$ & 600 \\
\multirow[t]{5}{*}{$DD^*\to D^*D^*$} 
& $\pi$ & $-3/2$ & $3/2$ & 600 \\
& $\eta$ & $1/6$ & $-1/6$ & 600 \\
& $\rho$ & $-3/2$ & $3/2$ & 600 \\
& $\omega$ & $1/2$ & $-1/2$ & 600 \\
\hline
\multirow[t]{5}{*}{$D^*D^*\to D^*D^*$} 
& $\sigma$ & $1$ & $-1$ & 600 \\
& $\pi$ & $-3/2$ & $3/2$ & 600 \\
& $\eta$ & $1/6$ & $-1/6$ & 600 \\
& $\rho$ & $-3/2$ & $3/2$ & 600 \\
& $\omega$ & $1/2$ & $-1/2$ & 600 \\
\hline
\hline
\end{tabular}
\end{table}

\setlength{\tabcolsep}{15pt}
\renewcommand{\arraystretch}{1.5}
\begin{table}[!htp]
\centering
\caption{
Values of the IS factors and reduced cutoff masses for each vertex and
isospin factor for the kernel amplitudes for $DD$, $DD^\ast$, and
$D^*D^*$ in the isoscalar ($I=0$) channel. 
} 
\label{tab:2}
\begin{tabular}{c|c|cc|c}
\hline
\hline
\multirow{2}{*}{Reaction} & \multirow{2}{*}{Exchange} & 
\multicolumn{2}{c|}{IS factor} & 
\multirow{2}{*}{$\Lambda_0$ [MeV]} \\
\cline{3-4}
& & $t$ channel & $u$ channel \\
\hline
\multirow[t]{5}{*}{$DD\to DD$} 
& $\sigma$ & $1$ & - & 600 \\
& $\rho$ & $1/2$ & $1/2$ & 600 \\
& $\omega$ & $1/2$ & $1/2$ & 600 \\
\multirow[t]{5}{*}{$DD\to DD^*$} 
& $\rho$ & $1/2$ & $1/2$ & 600 \\
& $\omega$ & $1/2$ & $1/2$ & 600 \\
\multirow[t]{5}{*}{$DD\to D^*D^*$} 
& $\pi$ & - & $1/2$ & 600 \\
& $\eta$ & - & $1/6$ & 600 \\
& $\rho$ & $1/2$ & $1/2$ & 600 \\
& $\omega$ & $1/2$ & $1/2$ & 600 \\
\hline
\multirow[t]{5}{*}{$DD^*\to DD^*$} 
& $\sigma$ & $1$ & - & 600 \\
& $\pi$ & - & $1/2$ & 600 \\
& $\eta$ & - & $1/6$ & 600 \\
& $\rho$ & $1/2$ & $1/2$ & 600 \\
& $\omega$ & $1/2$ & $1/2$ & 600 \\
\multirow[t]{5}{*}{$DD^*\to D^*D^*$} 
& $\pi$ & $1/2$ & $1/2$ & 600 \\
& $\eta$ & $1/6$ & $1/6$ & 600 \\
& $\rho$ & $1/2$ & $1/2$ & 600 \\
& $\omega$ & $1/2$ & $1/2$ & 600 \\
\hline
\multirow[t]{5}{*}{$D^*D^*\to D^*D^*$} 
& $\sigma$ & $1$ & - & 600 \\
& $\pi$ & $1/2$ & $1/2$ & 600 \\
& $\eta$ & $1/6$ & $1/6$ & 600 \\
& $\rho$ & $1/2$ & $1/2$ & 600 \\
& $\omega$ & $1/2$ & $1/2$ & 600 \\
\hline
\hline
\end{tabular}
\end{table}

The invariant amplitudes for pseudoscalar exchanges
are given by 
\begin{align}
\mathcal{M}_{P^*P\to P^*P}^{\mathcal{M}} &= 
-\mathrm{IS} F_{\mathcal{M}}^2 \frac{4g^2}{f_\pi^2} 
\frac{(\epsilon_1\cdot q)(\epsilon_3^*\cdot q)}
{q^2 - m_{\mathcal{M}}^2}, \cr
\mathcal{M}_{P^*P^*\to P^*P^*}^{\mathcal{M}} &= 
-\mathrm{IS} F_{\mathcal{M}}^2 \frac{g^2}{f_\pi^2} 
\varepsilon_{\mu\nu\alpha\beta}
\varepsilon^{\lambda\kappa\rho\sigma}\frac{q^\mu (p_1+p_3)^\beta
q_\lambda (p_2+p_4)_\sigma 
\epsilon_1^\nu\epsilon_2^\alpha\epsilon_3^{*\kappa}\epsilon_4^{*\rho}}
{q^2 - m_{\mathcal{M}}^2}.
\label{eq:27}
\end{align}
Those for vector-meson exchanges are written as
\begin{align}
\mathcal{M}_{PP\to PP}^V &= \mathrm{IS} F_V^2 \frac{\beta^2 g_V^2}{2}
\frac{(p_1+p_3)\cdot(p_2+p_4) - 
((p_1+p_3)\cdot q)((p_2+p_4)\cdot q)/m_V^2}{q^2 -m_V^2}, \cr
\mathcal{M}_{P^*P\to P^*P}^V &= \mathrm{IS} F_V^2 \lambda\beta g_V^2 
\varepsilon_{\beta\alpha\mu\nu}(p_1+p_3)^\beta q^\mu
\frac{g^{\nu\sigma} - q^\nu q^\sigma/m_V^2}{q^2 - m_V^2}
(p_2+p_4)_\sigma \epsilon_1^\alpha\epsilon_3^*, \cr
\mathcal{M}_{P^*P^*\to P^*P^*}^V &= -\mathrm{IS} F_V^2 g_{P^*P^*V}^2 
\frac{1}{q^2 - m_V^2} 
\left[(p_1+k_1)\cdot(p_2+k_2)(\epsilon_1\cdot\epsilon_3^*)
(\epsilon_2\cdot\epsilon_4^*) \right. \nonumber \\
&\quad \left.-(\epsilon_1\cdot\epsilon_4^*)
(\epsilon_2\cdot\epsilon_3^*)
  ((p_1-k_1)\cdot(p_2-k_2)) \right].
\label{eq:28}  
\end{align}

\section{Results and discussion}
While the parity of the $T_{cc}(3875)$ is known to be positive, we
will show that tetraquark states with both positive and negative
parities are dynamically generated. Since we are mainly concerned with
their production mechanism, we do not fit the experimental data by
adjusting parameters such as coupling constants and reduced cutoff
masses. We begin by discussing the partial-wave kernel
amplitudes. Interestingly, three bound states emerge even at the level
of single channels. Once coupled channels are introduced, the
positions of the bound states shift either along the real axis or into
the complex plane, and an additional resonance appears above the
$D^*D^*$ threshold with negative parity. This will be discussed in the
subsection on transition amplitudes. We then examine the pole
positions of the tetraquark resonances and their coupling strengths to
each channel. Finally, we investigate the uncertainties that arise
from varying the values of the reduced cutoff mass. 
\subsection{Kernel amplitudes}
The kernel amplitudes are determined by the Feynman invariant
amplitudes given in Eqs.~\eqref{eq:27} and~\eqref{eq:28}. To generate
resonances, it is essential to have attractive interactions in the
kernel amplitudes. By examining the partial-wave kernel amplitudes, we
can understand how the tetraquark resonances are generated. 

\subsubsection{Positive parity}
\begin{figure}[!htp] 
\centering
\includegraphics[scale=0.57]{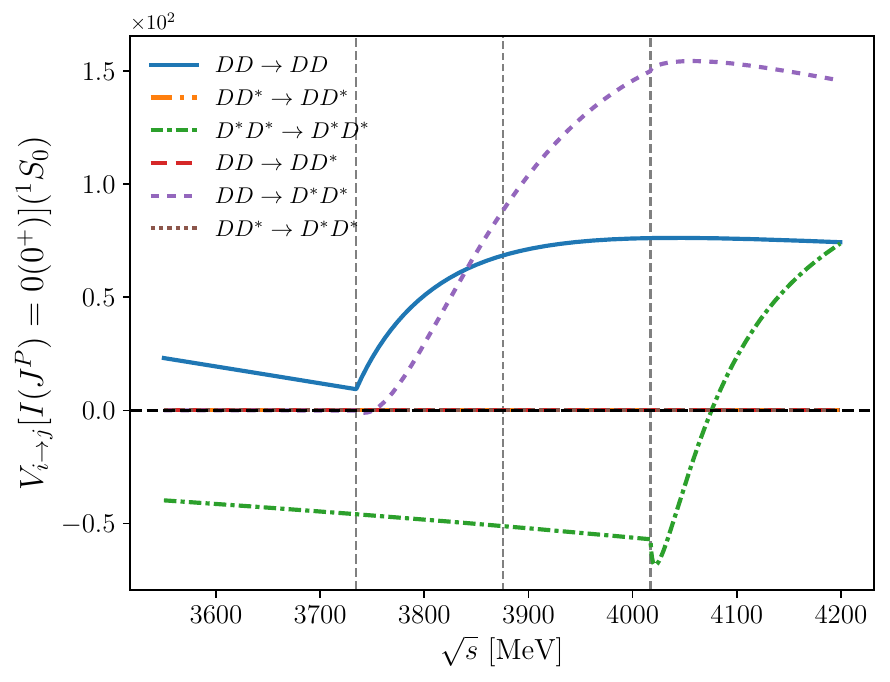}
\hspace{1em}
\includegraphics[scale=0.57]{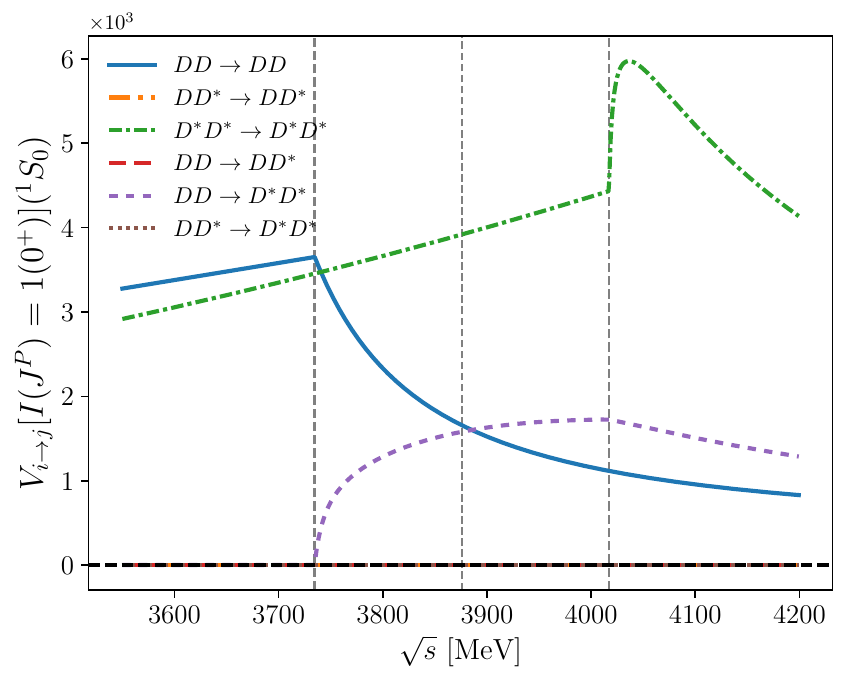}
\caption{Scalar kernel amplitudes as functions of $\sqrt{s}$. The left
  panel depicts the results for those with $I=0$ and $J=0$, whereas
  the right panel shows the results for those with $I=1$ and $J=0$.}  
\label{fig:4}
\end{figure}
In Fig.~\ref{fig:4}, we present the numerical results
for the partial-wave kernel amplitudes $\mathcal{V}_{ij}^J$
for the isoscalar-scalar channel ($I=0$, $J=0$). Since the 
contributions from partial-wave amplitudes higher than the $P$-wave 
turn out to be negligible, we focus on the scalar ($S$-wave) kernel 
amplitudes for the positive-parity channels. Note that the $P$-wave is not
considered because of the parity. In the isoscalar-scalar
channel shown in the left panel of Fig.~\ref{fig:4}, the amplitudes
involving the $DD^*$ state vanish. The elastic $DD$ and $DD \to D^*D^*$
kernel amplitudes are positive over the entire range of the
cm energy, so they cannot generate any resonance. Although
the elastic $D^*D^*$ amplitude is negative up to $\sqrt{s} \approx
4.1$ GeV, it is almost canceled out by the other two amplitudes. This
implies that resonances do not appear in the isoscalar-scalar channel. 
For the scalar-isovector channel ($J=0$, $I=1$), depicted in the
right panel of Fig.~\ref{fig:4}, we draw the same conclusion: 
isovector-scalar resonances are also absent, since all the nonvanishing
amplitudes are positive.

\begin{figure}[!htp]
\centering
\includegraphics[scale=0.57]{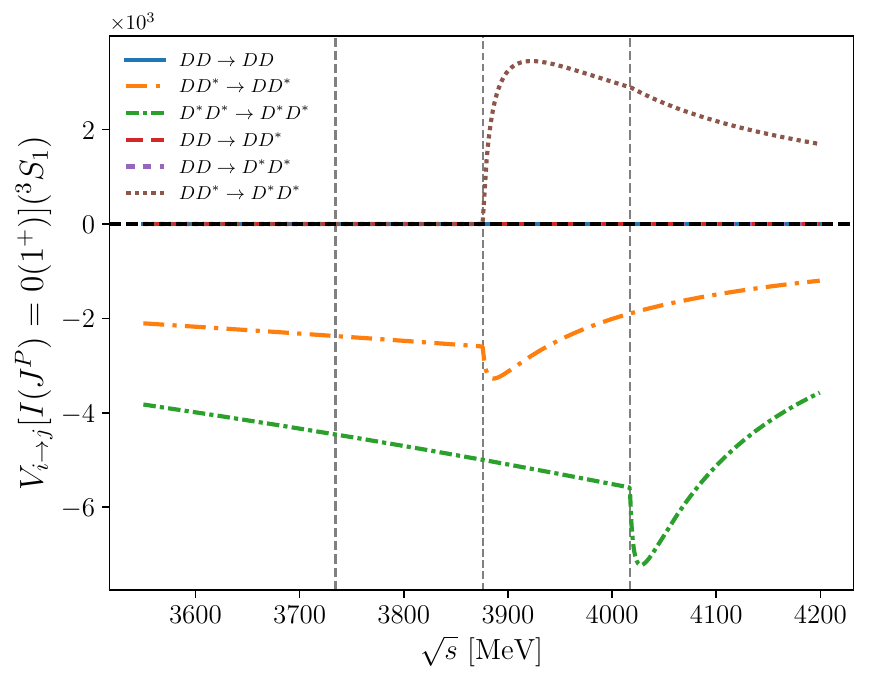}
\hspace{1em}
\includegraphics[scale=0.57]{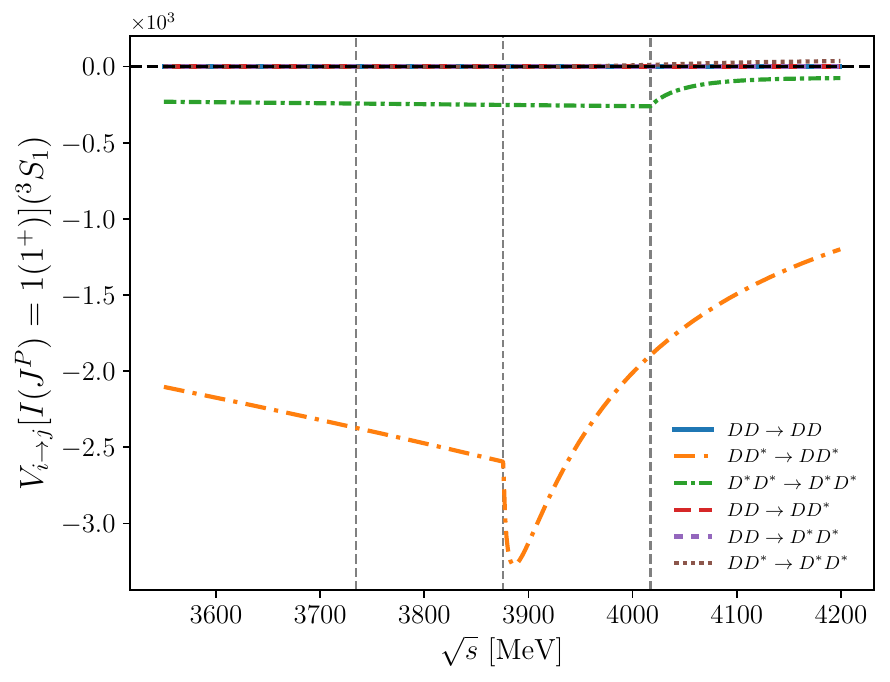}
\caption{
Axial-vector kernel amplitudes as functions of
$\sqrt{s}$. The left panel depicts the results for those with $I=0$
and $J=1$, whereas the right panel shows the results for those with
$I=1$ and $J=1$.} 
\label{fig:5}
\end{figure}
In Fig.~\ref{fig:5}, on the other hand, the kernel amplitudes in the $J=1$
channel provide strong attractions, which are crucial for generating
tetraquark resonances. The left panel of Fig.~\ref{fig:5}
shows that the elastic isoscalar--axial-vector $DD^*$ and
$D^*D^*$ amplitudes are negative, whereas the $DD^* \to D^*D^*$ amplitude is
repulsive throughout the entire range of the cm
energy. This leads to the formation of two bound states in the
isoscalar--axial-vector channel. In the right panel of
Fig.~\ref{fig:5}, we present the isovector--axial-vector kernel
amplitudes. While the overall behavior is similar to that of the
isoscalar--axial-vector channel, the elastic $DD^*$ amplitude becomes
the most dominant. Therefore, a bound state also arises in the $DD^*$
channel. It is noteworthy that the $DD^* \to D^*D^*$ transition
amplitude nearly vanishes. This results from the fact that the
contributions from $\rho$- and $\omega$-exchange almost cancel each
other due to the different signs of the IS factor. 

\subsubsection{Negative parity}
\begin{figure}[!htp] 
\centering
\includegraphics[scale=0.57]{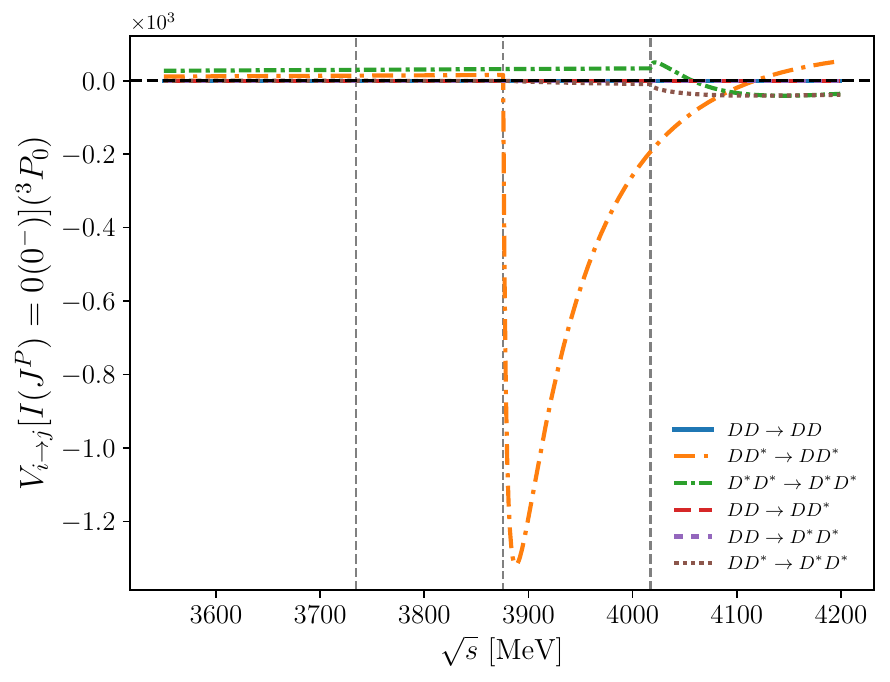}
\hspace{1em}
\includegraphics[scale=0.57]{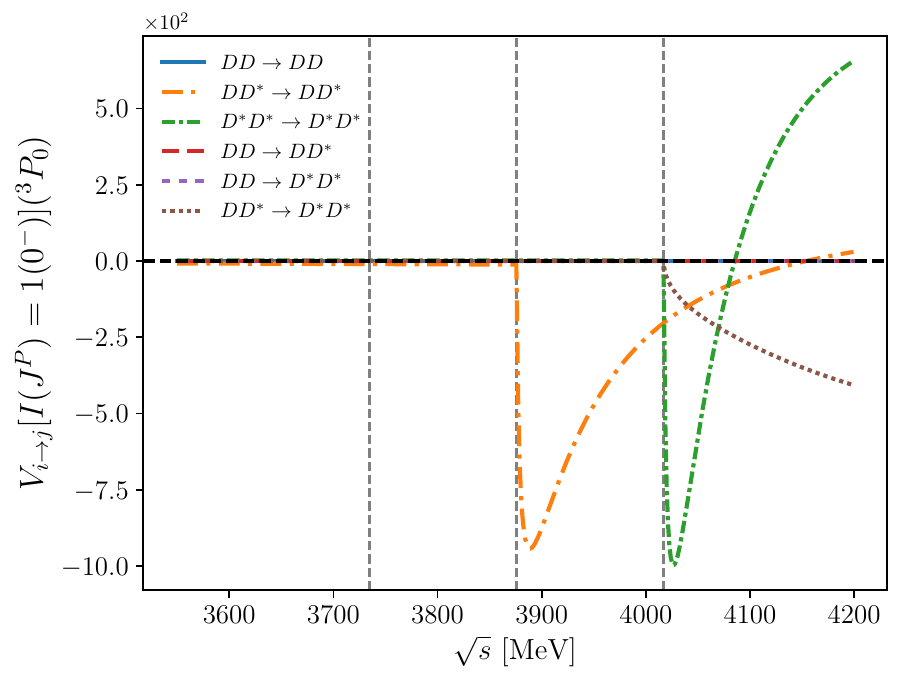}
\caption{
  Pseudoscalar kernel amplitudes as functions of
  $\sqrt{s}$. The left panel depicts the results for those with $I=0$
  and $J=0$, whereas the right panel shows the results for those with
  $I=1$ and $J=0$.
}
\label{fig:6}
\end{figure}
The ${}^3P_0$ channel with $J=0$, as well as the ${}^1P_1$ and
${}^3P_1$ channels with $J=1$, are responsible for the negative-parity
resonances. In Fig.~\ref{fig:6}, the $P$-wave kernel
amplitudes are illustrated in the $LSJ$ basis. In the $J=0$ channel,
the conservation of total angular momentum requires $S=1$, so that the
elastic $DD$ kernel amplitude must vanish, as shown in the left panel
of Fig.~\ref{fig:6}. The elastic $DD^*$ kernel amplitude dominates over
all other amplitudes and provides a strong attraction. 
Similar features are observed in the $J=1$ channel, where the elastic
$DD^*$ and $D^*D^*$ amplitudes dominate over the others, as
demonstrated in the right panel of Fig.~\ref{fig:6}. Although the
corresponding kernel amplitudes exhibit negative bump structures,
bound states do not appear in the negative-parity case.

\begin{figure}[!htp] 
\centering
\includegraphics[scale=0.57]{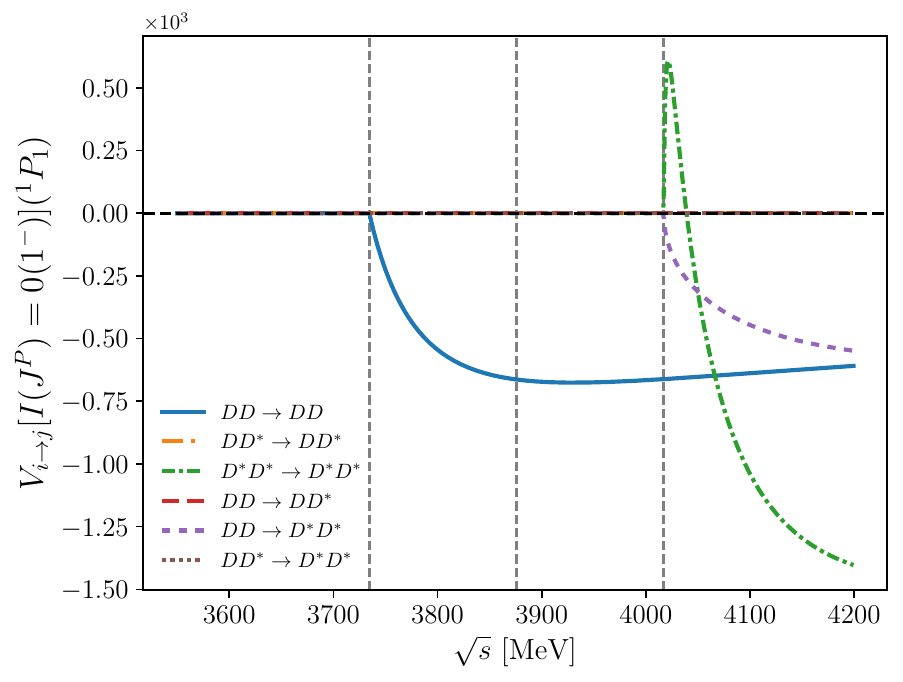}
\hspace{1em}
\includegraphics[scale=0.57]{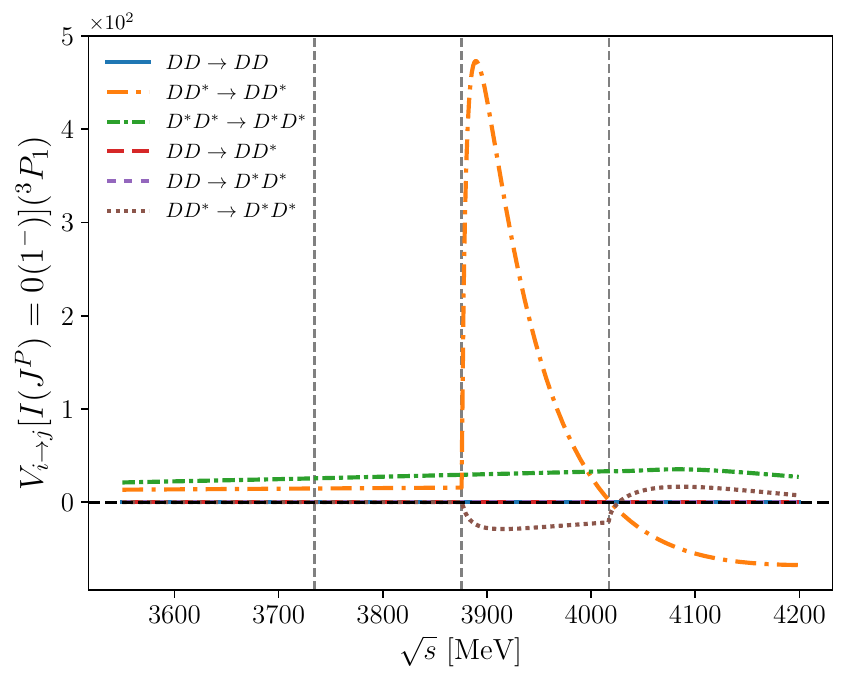}
\includegraphics[scale=0.57]{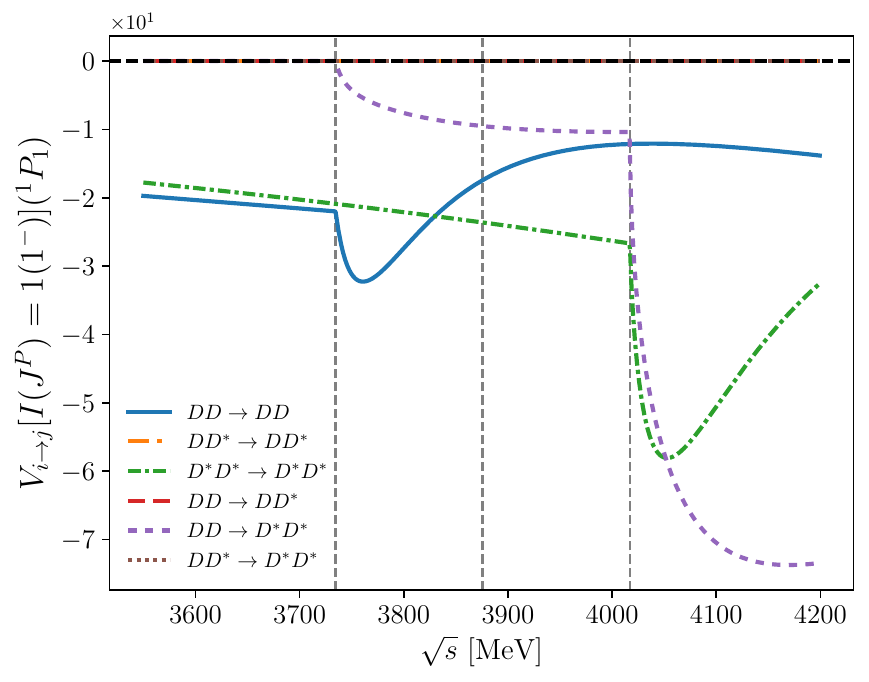}
\hspace{1em}
\includegraphics[scale=0.57]{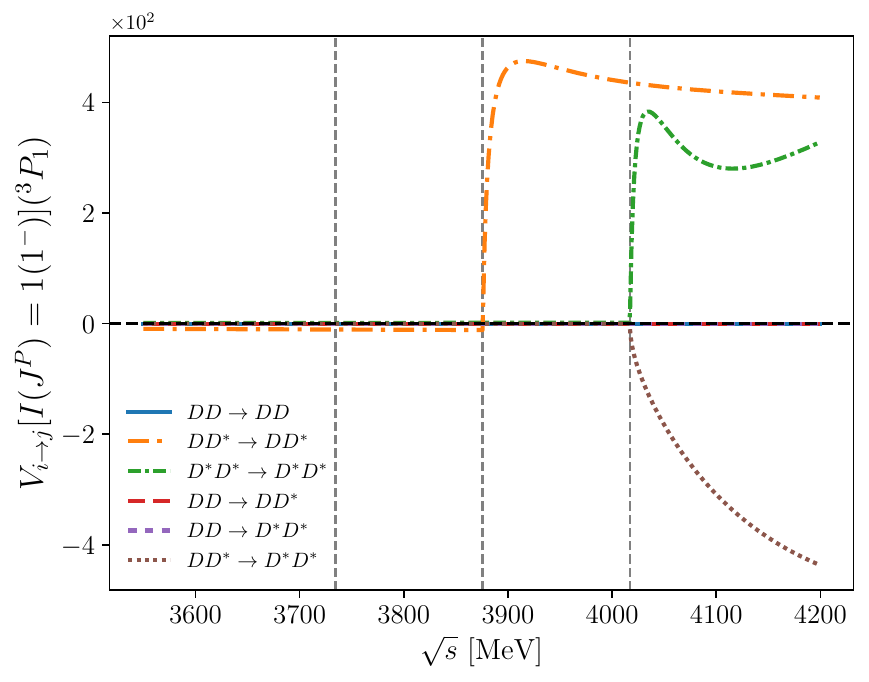}
\caption{
  Vector kernel amplitudes as functions of
  $\sqrt{s}$. The left panel depicts the results for those with $I=0$
  and $J=1$, whereas the right panel shows the results for those with
  $I=1$ and $J=1$.
}
\label{fig:7}
\end{figure}
The left and right panels of Fig.~\ref{fig:7} show the results for the
$P$-wave ${}^1P_1$ and ${}^3P_1$ kernel amplitudes with $J=1$,
respectively. The isoscalar ${}^1P_1$ amplitudes, displayed in the
upper-left panel of Fig.~\ref{fig:7}, are much larger than the other
$P$-wave amplitudes. While they are mostly attractive, they do not
lead to the formation of bound states. Therefore, the negative-parity
kernel amplitudes are conclusively not strong enough to generate
tetraquark bound states. 

\subsection{Transition amplitudes}
We are now in a position to discuss the results for the 
transition amplitudes. The partial-wave transition amplitudes are
obtained by solving Eq.~\eqref{eq:7}, which indicates that all the
channels are coupled to each other. To understand the behavior of the
coupled transition amplitudes, it is worthwhile to examine each single
channel separately. Thus, we first scrutinize a single channel by switching
off the other channels in Eq.~\eqref{eq:7}, and then continue to
investigate the features of the fully coupled transition amplitudes.
When we consider single channels with $J=0$, we do not find any bound
states. For $J=1$, we observe three bound states. It is important to
note that only when all channels are coupled does one of them become a
resonance in the complex plane. The other two bound states are shifted
due to the coupled-channel effects.

\subsubsection{Positive parity}
\begin{figure}[!htp]
\centering
\includegraphics[scale=0.57]{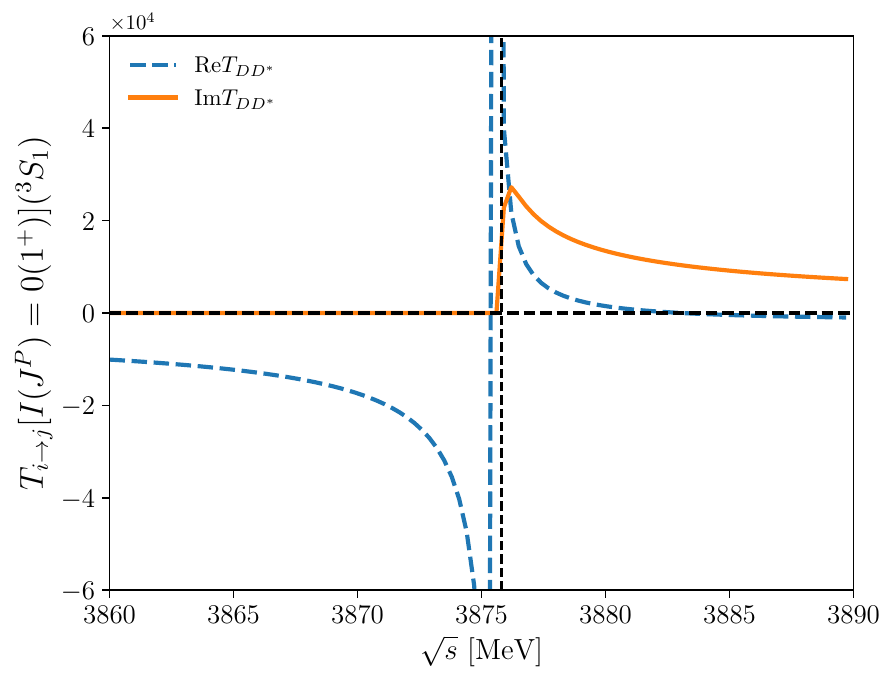}
\hspace{1em}
\includegraphics[scale=0.57]{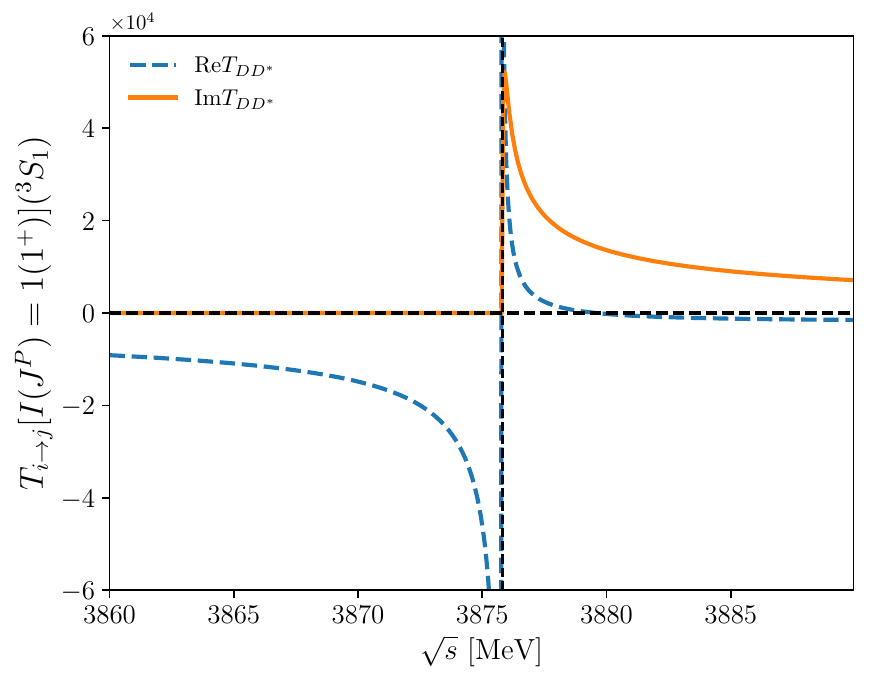}
\caption{
  Single-channel transition amplitudes for elastic $DD^*$
  scattering in the ${}^3S_1$ channels with isospin $I=0$ (left panel)
  and $I=1$ (right panel) as functions of the cm energy. The vertical
  dashed lines indicate the $DD^*$ threshold.
}
\label{fig:8}
\end{figure}
We first examine the elastic $DD^*$ single channel. 
In the left and right panels of Fig.~\ref{fig:8}, we clearly see that 
the $S$-wave amplitudes $T_{DD^* \to DD^*} (I=0, J=1)$ and $T_{DD^*
\to DD^*} (I=1, J=1)$ exhibit bound-state structures near the $DD^*$ 
mass threshold at $\sqrt{s} = 3875.81$ MeV. As shown in
Fig.~\ref{fig:5}, the $DD^*$ kernel amplitudes provide strong
attractions, which lead to the generation of two bound states
slightly below the threshold in the ${}^3S_1$ channels, corresponding
to different isospins. We also predict the isovector bound state, as
demonstrated in the right panel of Fig.~\ref{fig:8}.

\begin{figure}[!htp]
\centering
\includegraphics[scale=0.57]{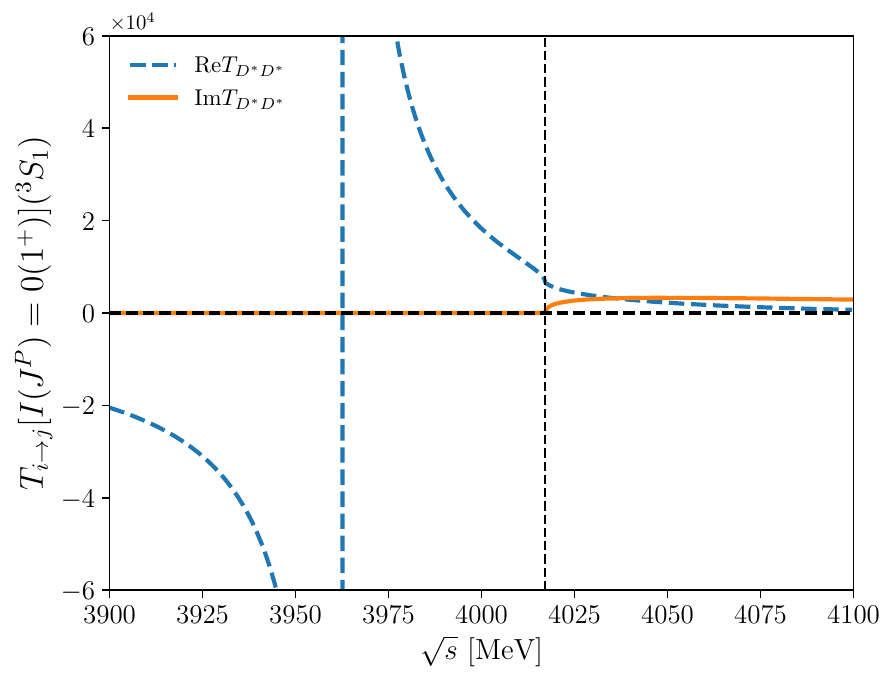}
\includegraphics[scale=0.57]{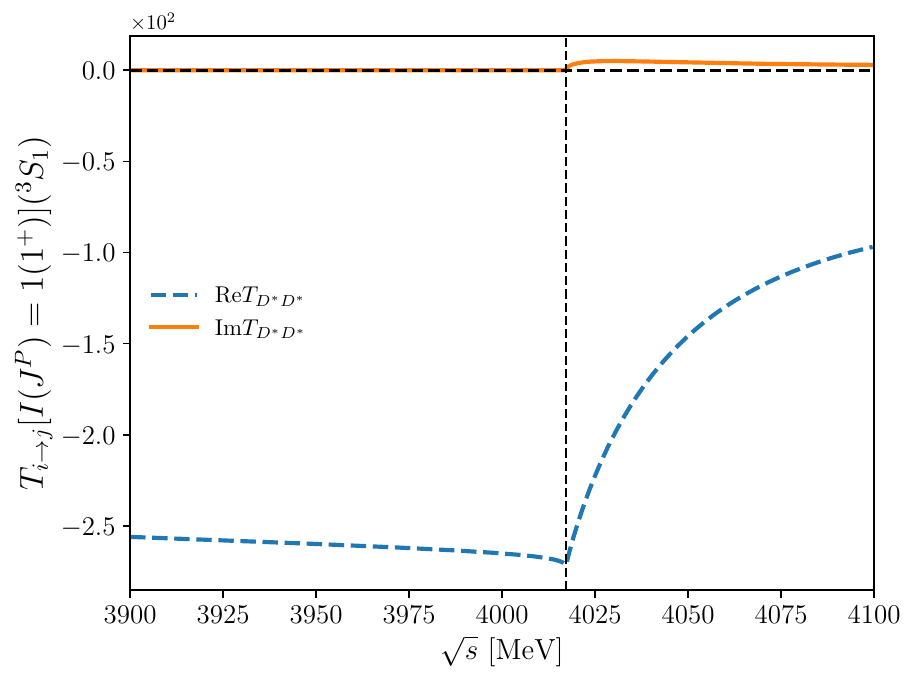}
\caption{
  Single-channel transition amplitudes for elastic $D^*D^*$
  scattering in the ${}^3S_1$ channels with isospin $I=0$ (left panel)
  and $I=1$ (right panel) as functions of the cm energy. The vertical
  dashed lines indicate the $D^*D^*$ threshold.
}
\label{fig:9}
\end{figure}
Figure~\ref{fig:9} presents the results for the single-channel $D^*D^*
\to D^*D^*$ transition amplitudes. The left panel shows the results
for the isoscalar channel ($I=0$), whereas the right one displays
those for the isovector channel ($I=1$). In the isoscalar case, a deep
bound state is generated significantly below the $D^*D^*$ mass 
threshold. The corresponding pole is located at $\sqrt{s} =
3962.55\ \mathrm{MeV}$, implying a binding energy of approximately
$54.57$ MeV relative to the threshold. This strong attraction in the
$D^*D^*$ elastic channel is consistent with the kernel amplitude
results shown in Fig.~\ref{fig:5} for $J=1$ and $I=0$. In 
contrast, the isovector channel does not exhibit any bound-state
structure near the $D^*D^*$ threshold. Thus, no bound state or
resonance is found in the $D^*D^*$ channel with $I=1$ within the
present framework.

\begin{figure}[!htp]
    \centering
    \includegraphics[scale=0.57]{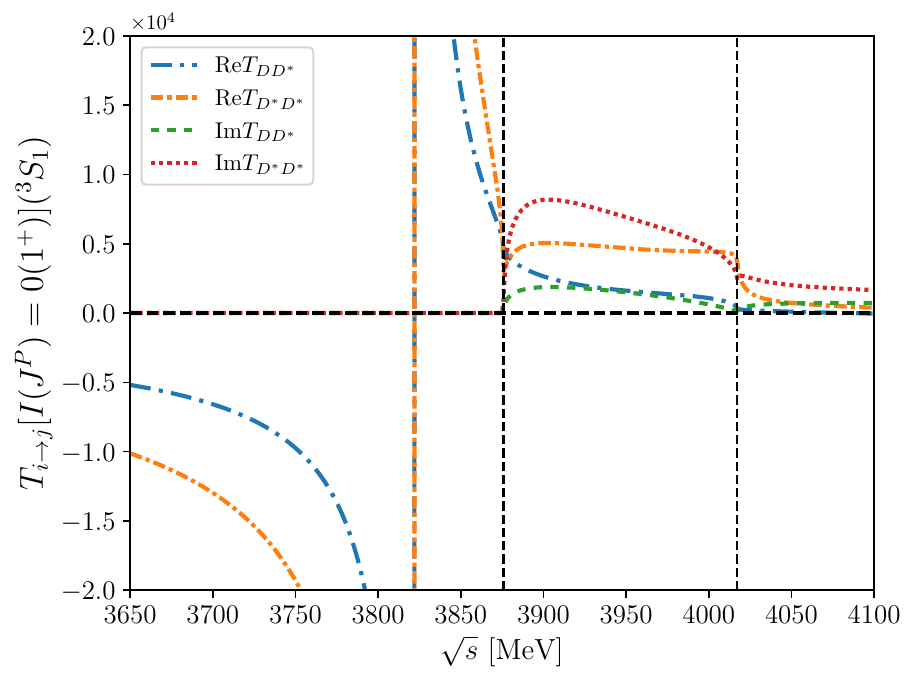}
    \includegraphics[scale=0.57]{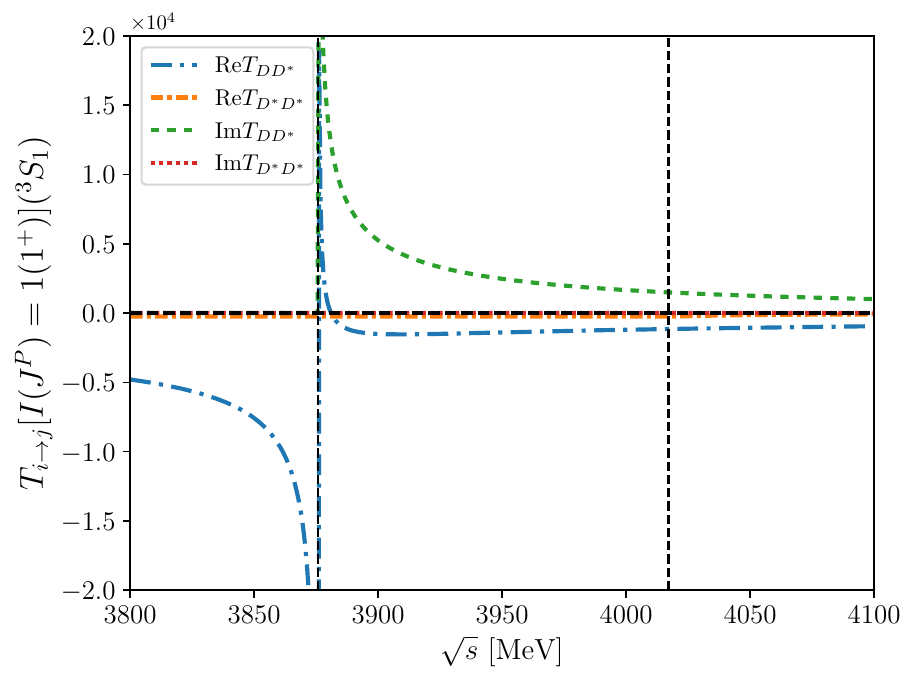}
    \caption{
      Transition amplitudes for the elastic $DD^\ast$ and
      $D^\ast D^\ast$ interactions in the ${}^3S_1$ channels with
      isospin $I=0$ (left panel) and $I=1$ (right panel) as functions
      of the cm energy, with all channels switched on. The
      vertical dashed lines indicate the $DD^*$ and $D^*D^*$
      thresholds. 
}
    \label{fig:10}
  \end{figure}
We now scrutinize the coupled-channel effects on the elastic $DD^*$
and $D^*D^*$ interactions. Figure~\ref{fig:10} shows the
coupled-channel transition amplitudes in the ${}^3S_1$ channels
with isospin $I=0$ (left panel) and $I=1$ (right panel) as functions 
of the cm energy. In the isoscalar channel, a pole appears
as a result of the combined attractions in the $DD^*$ and $D^*D^*$
channels. The attractive kernel amplitudes compete with the repulsive
one in the $DD^* \to D^*D^*$ transition, as shown in
Fig.~\ref{fig:5}. Consequently, the pole becomes more deeply bound 
than in the single-channel case, with a binding energy of
approximately $58$~MeV. Since the $DD$ threshold is absent in this
$S$-wave channel, there is no open channel below the $DD^*$
threshold. As a result, we find a bound state instead of a
resonance. Had we considered three-body channels such as the $DD\pi$
channel, the bound state would have moved to the complex plane. 

Interestingly, the bound state that existed in the $D^*D^*$ channel,
initially located between the $DD^*$ and $D^*D^*$ thresholds,
disappears in the fully coupled transition amplitude as a  
result of the $DD^*$ coupled-channel effect. The isovector bound
state, on the other hand, remains largely unaffected by the $D^*D^*$
potential and appears close to the $DD^*$ threshold, with a binding
energy of about $10$~keV. 
Turning on the full coupled-channel effects, we observe that the bound
state around 3960 MeV in the $D^*D^*$ single channel (see
Fig.~\ref{fig:9}) evolves into a resonance, with the pole
position located at 4102 MeV and a decay width of $166$ MeV. This
indicates that the effects of the $DD^* \to DD^*$ and $DD^* \to D^*D^*$
channels are crucial in predicting this resonance state.

\subsubsection{Negative parity}
\begin{figure}[!htp] 
\centering
\includegraphics[scale=0.57]{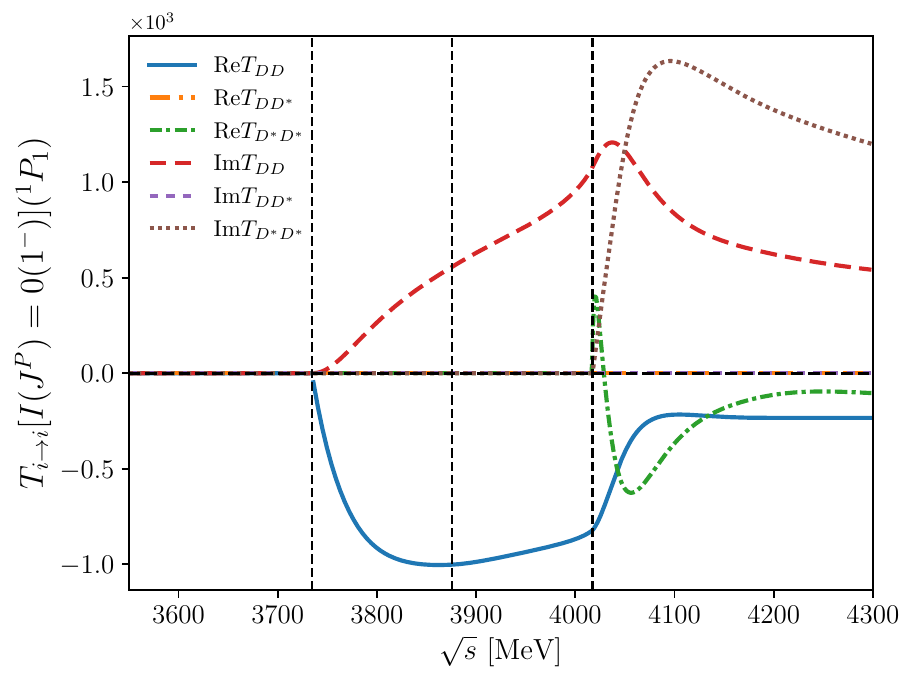}
\includegraphics[scale=0.57]{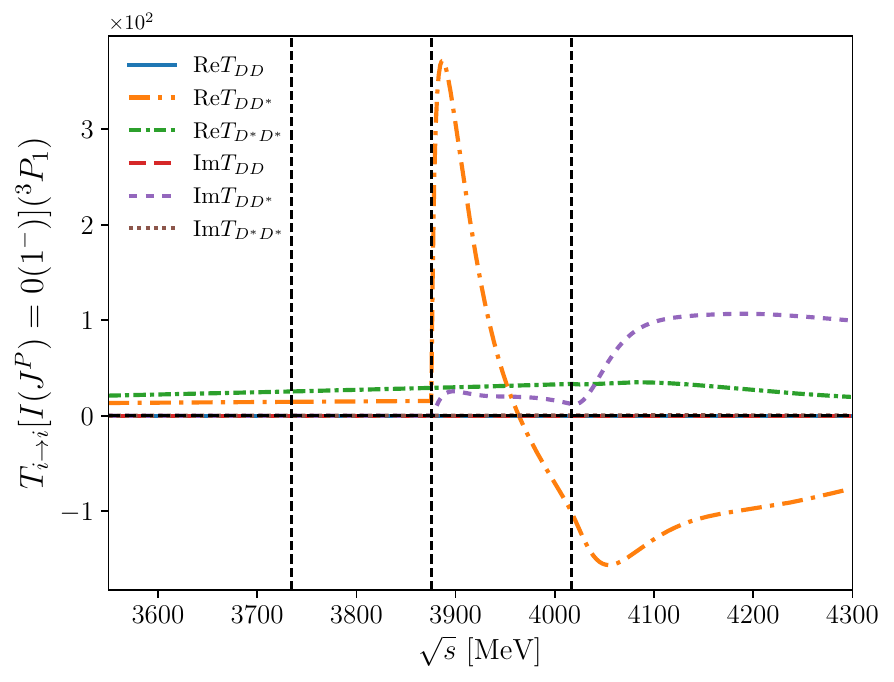}
\caption{
Transition amplitudes for the elastic $DD$, $DD^\ast$, and
$D^\ast D^\ast$ interactions in the ${}^1P_1$ channel (left
panel) and ${}^3P_1$ channel with $I=0$ (right panel) as
functions of the cm energy, with all channels
switched on. The vertical dashed lines indicate the
$DD$, $DD^*$, and $D^*D^*$ thresholds.
}
\label{fig:11}
\end{figure}
As shown in the left panel of Fig.~\ref{fig:7}, the kernel amplitudes
for the ${}^1P_1$ channels are mostly attractive. This leads us
to expect bound states even in the single-channel transition
amplitudes with negative parity. However, none of the single-channel
$T$ amplitudes exhibit any indication of such states. When we consider
the full coupled-channel case, however, a resonance emerges. The left
and right panels of Fig.~\ref{fig:11} display the transition
amplitudes for the ${}^1P_1$ and ${}^3P_1$ channels with $I=0$,
respectively. The results shown in Fig.~\ref{fig:11} indicate that the
resonance with negative parity consists of both the isoscalar
${}^1P_1$ and ${}^3P_1$ channels. Thus, this negative-parity
tetraquark state emerges as a $P$-wave mixed state, meaning that it
cannot be identified as a molecular state. 

\subsection{Pole positions of the tetraquark states and  their
  coupling strengths}
We can examine the pole positions of the tetraquark states by scanning
the transition amplitudes in the complex energy plane. The resulting
pole positions and coupling strengths to each channel
are listed in Table~\ref{tab:1}. Note that the units of the coupling
strengths are given in GeV. They can be extracted from the transition
amplitudes in the vicinity of the pole position 
$\sqrt{s_R}$ as follows: 
\begin{align}
  \label{eq:29}
T_{a\to a}(\sqrt{s}) = 4\pi\frac{g_a^2}{s-s_R},
\end{align}
where the residue $g_a$ ($g_b$) is the coupling strength corresponding
to channel $a$ ($b$).  

\setlength{\tabcolsep}{15pt}
\renewcommand{\arraystretch}{1.5}
\begin{table}[htbp]
\caption{Coupling strengths $g_i$ for exotic mesons $T_{cc}$, labeled
  by the real part of the pole positions.} 
  \centering
\begin{tabular}{l|c|c|c|c}
\hline\hline
$I(J^P)$ & \multicolumn{2}{c|}{$0(1^+)$} & $1(1^+)$ & $0(1^-)$ \\
\hline
State & $T_{cc}^{I=0}(3823)$ & $T_{cc}^{I=0}(4102)$ &
 $T_{cc}^{I=1}(3875)$ & $T_{cc}^{I=0}(4090)$ \\
\hline
$\sqrt{s_R}[\text{MeV}]$ & 3823.058 & $4102.326 - i83.044$ &
$3875.677$ & $4089.67 - i41.88$ \\
\hline
$g_{DD({}^1P_1)}$        & $-$ & $-$ & $-$ & $4.834-i2.214$ \\
$g_{DD^*({}^3S_1)}$      & $17.95$ & $1.310+i1.045$ & $3.276$ & $-$ \\
$g_{DD^*({}^3D_1)}$      & $0.113$ & $0.106+i0.2113$ & $1.342\times10^{-3}$ & $-$ \\
$g_{DD^*({}^3P_1)}$      & $-$ & $-$ & $-$ & $0.289+i2.035$ \\
$g_{D^*D^*({}^3S_1)}$    & $24.97$ & $0.299+i1.310$ & $2.147\times10^{-2}$ & $-$ \\
$g_{D^*D^*({}^3D_1)}$    & $0.589$ & $0.148+i0.5503$ & $3.794\times10^{-4}$ & $-$ \\
$g_{D^*D^*({}^5D_1)}$    & $3.325\times10^{-4}$ & $(4.800+i2.246)\times10^{-3}$ & $4.446\times10^{-4}$ & $-$ \\
$g_{D^*D^*({}^1P_1)}$    & $-$ & $-$ & $-$ & $2.331-i7.371$ \\
$g_{D^*D^*({}^3P_1)}$    & $-$ & $-$ & $-$ & $(2.624+i1.751)\times10^{-3}$ \\
$g_{D^*D^*({}^5P_1)}$    & $-$ & $-$ & $-$ & $4.110-i1.168$ \\
$g_{D^*D^*({}^5F_1)}$    & $-$ & $-$ & $-$ & $1.707-i0.732$ \\
\hline\hline
\end{tabular}
\label{tab:3}
\end{table}
As discussed above, we can identify four states with total spin $J =
1$: three axial-vector tetraquark states with positive parity and one
vector tetraquark with negative parity. In the isoscalar channel, we
find a $DD^*$ bound state at $3823.058~\mathrm{MeV}$ on the real
axis. Its coupling strength to the $D^*D^*$ channel is the most
dominant, followed by that to the $DD^*$ channel. The coupling
strengths to all other channels are either negligible or vanish. Thus,
inclusion of the $D^*D^*$ channel is essential to form this bound
state. The second isoscalar axial-vector tetraquark state appears as a
resonance located at $\sqrt{s_R} = (4102.326 -
i\,93.044)~\mathrm{MeV}$. We find that its coupling strengths to the
$DD^*$ and $D^*D^*$ channels dominate over all others, similar to the
case of $T_{cc}(3823)$. Since it lies above the $D^*D^*$ threshold, it
can decay into both the $DD^*$ and $D^*D^*$ channels. 

The isovector axial-vector tetraquark state also emerges as a bound
state on the real axis at $3875.677~\mathrm{MeV}$. It lies almost
exactly at the $DD^*$ threshold, with a binding energy given by the
small value $\delta m = m_{DD^*} - m_R = 0.0328~\mathrm{MeV}$,
indicating a loosely bound state. This state is in good agreement
with the existing $T_{cc}(3875)$ state reported by the LHCb
Collaboration~\cite{LHCb:2021vvq}. Though the quantum numbers of the
$T_{cc}(3875)$ have not been clearly identified, we consider it
favorably as an isovector doubly charmed tetraquark state. It is
remarkable to observe that its coupling strength to the $DD^*$ channel
is the most dominant. This implies that the $T_{cc}(3875)$ can be
regarded as a $DD^*$ molecular state. 

As discussed previously, an interesting feature of the present work is
that we also predict an isoscalar tetraquark resonance with negative
parity in addition to the positive-parity ones. It appears at
approximately $5~\mathrm{MeV}$ above the $D^*D^*$ threshold, with a
total width of $93.85~\mathrm{MeV}$. It is noteworthy that the
contributions from higher partial waves such as ${}^5P_1$ and
${}^5F_1$ are sizable in the coupling strengths to the $D^*D^*$
channel. Thus, by examining the coupling strengths to various
channels, we understand that this negative-parity tetraquark state
emerges as the most dynamically generated one. It may possibly be
regarded as a genuine tetraquark state. 

\subsection{Uncertainty of the reduced cutoff mass}
Though we do not fit the values of the reduced cutoff mass
$\Lambda_0$, it is important to estimate the uncertainty arising from
variations in $\Lambda_0$. We examine the four different tetraquark
states produced in the present framework with $\Lambda_0$ varied from
$500~\mathrm{MeV}$ to $700~\mathrm{MeV}$. The central value,
$\Lambda_0 = 600~\mathrm{MeV}$, is adopted in the current work.

\setlength{\tabcolsep}{15pt}
\renewcommand{\arraystretch}{1.5}
\begin{table}[htbp]
\caption{Dependence of the pole positions on the reduced cutoff mass
$\Lambda_0 = \Lambda - m_\mathrm{ex}$.}  
\centering
\begin{tabular}{l|c|c|c}
\hline\hline
\multirow{2}{*}{Tetraquark states ($J^P$)} &
 \multicolumn{3}{c}{$\Lambda_0$~[MeV]}  \\ 
\cline{2-4}
 & $500$ & $600$ & $700$ \\
\hline
$T_{cc}^{I=0}(3823)\, (1^+)$ & $3867.488$ & $3823.06$ & $3750.0$ \\
$T_{cc}^{I=0}(4102)\,(1^+)$ & $4103.260-i82.05$ & $4102.326-i83.044$ & $4102.346-i84.15$ \\
$T_{cc}^{I=1}(3875)\,(1^+)$ & Cusp & $3875.677$ & $3871.78$ \\
$T_{cc}^{I=0}(4090)\,(1^-)$ & $4063.66-i67.47$ & $4089.67-i41.88$ & Virtual state \\
\hline\hline
\end{tabular}
\label{tab:4}
\end{table}
Table~\ref{tab:4} shows how the pole positions change as $\Lambda_0$
varies. When we take $\Lambda_0 = 500~\mathrm{MeV}$, which is
considered to be phenomenologically rather soft, $T_{cc}(3875)$
reduces to a cusp structure. The mass of $T_{cc}(3823)$ increases by
about $45~\mathrm{MeV}$. The $T_{cc}(4102)$ remains almost unchanged.
On the other hand, the mass of the negative-parity $T_{cc}(4090)$ is
reduced by about $25~\mathrm{MeV}$, while its width increases.
When we use $\Lambda_0 = 700~\mathrm{MeV}$, the mass of
$T_{cc}(3875)$ remains nearly stable, whereas that of $T_{cc}(3823)$
continues to decrease. The $T_{cc}(4102)$ remains nearly unchanged.
Interestingly, the negative-parity $T_{cc}(4090)$ becomes a virtual
state. As long as $\Lambda_0$ is kept within the range of
$(600-700)~\mathrm{MeV}$, the most significant tetraquark state
$T_{cc}(3875)$ remains almost robust.
\section{Summary and conclusions}
In the present work, we have investigated the production mechanism of
the doubly charmed tetraquark states within a coupled-channel
framework. The kernel amplitudes are constructed by using an effective
Lagrangian that respects heavy quark symmetry, chiral symmetry, SU(3)
flavor symmetry, and hidden local symmetry. The scattering equation is
solved in the Blankenbecler–Sugar (BbS) scheme, which is a
three-dimensional reduction of the Bethe–Salpeter equation. We have
examined the pole structures of the resulting transition amplitudes
and the corresponding coupling strengths to different channels. The
cutoff masses are restricted to a reduced form $\Lambda_0 = \Lambda -
m_{\mathrm{ex}}$, in order to minimize model dependence. 

By analyzing the partial-wave kernel amplitudes, we have shown that
the isoscalar--axial-vector- and isovector--axial-vector exhibit 
strong attractions in the $S$-wave, and give rise to three bound
states. On the other hand, the negative-parity resonances appear as
$P$-wave mixed states, which do not qualify as hadronic molecules. We
have demonstrated that the isoscalar--axial-vector channel produces a
bound state at $3823$~MeV and a resonance at $(4102 - i\,93)$~MeV,
while the vector-isovector channel generates a
looesly bound state at $3875$~MeV. These results support the
interpretation of the observed $T_{cc}(3875)^+$ state as a $DD^*$
molecular state in the isovector channel. Moreover, we have predicted
a negative-parity isoscalar resonance around $4090$~MeV. Finally, we
have investigated the dependence of the pole positions on the reduced
cutoff mass $\Lambda_0$, and found that the main features of the
$T_{cc}(3875)$ remain robust within the range $\Lambda_0
=(600-700)$~MeV. 

\begin{acknowledgments}
The present work was supported by Inha University Grant
(Grant-No. 75463-1) in 2025.  
\end{acknowledgments}

\bibliography{DDTcc}
\bibliographystyle{apsrev4-1}

\end{document}